\documentclass[twocolumn, aps]{revtex4}

\usepackage{mathbbol}              

\usepackage{graphicx}

\usepackage{ulem}

\usepackage{color}
\usepackage{hyperref}
\usepackage[usenames,dvipsnames]{xcolor}
\usepackage{breakurl}
\usepackage{amsmath, amsthm, amssymb}

\begin{document}

\title{\Large Storage and control of optical photons using Rydberg polaritons}

\date{\today}
\author{D. Maxwell$^{\textrm{1\hyperref[con1]{*}}}$}
\author{D. J. Szwer$^{\textrm{1}}$}
\author{D. P. Barato$^{\textrm{1}}$}
\author{H. Busche$^{\textrm{1}}$}
\author{J. D. Pritchard$^{\textrm{1}\dagger}$}
\author{A. Gauguet$^{\textrm{1}\ddagger}$}
\author{K. J. Weatherill$^{\textrm{1}}$}
\author{M. P. A. Jones$^{\textrm{1}}$}
\author{C. S. Adams$^{\textrm{1\hyperref[con2]{*}}}$}
\affiliation{$^{\textrm{1}}$ Joint Quantum Centre (JQC) Durham-Newcastle,
Department of Physics,
Durham University,
Rochester Building,
South Road,
Durham DH1 3LE,
England.
\\{$^{\dagger}$Now at: Department of Physics,
University of Strathclyde,
107 Rottenrow East,
Glasgow G4 0NG, Scotland.}
\\{$^{\ddagger}$Now at: Laboratoire Collisions Agr\'egats R\'eactivit\'e,
Universit\'e Paul Sabatier,
B\^at.\ 3R1b4,
118 route de Narbonne,
31062 Toulouse Cedex 09, France.}}

\begin{abstract}  
We use a microwave field to control the quantum state of optical photons stored in a cold atomic cloud. The photons are stored in highly excited collective states (Rydberg polaritons) enabling both fast qubit rotations and control of photon-photon interactions. Through the collective read-out of these pseudo-spin rotations it is shown that the microwave field modifies the long-range interactions between polaritons. This technique provides a powerful interface between the microwave and optical domains, with applications in quantum simulations of spin liquids, quantum metrology and quantum networks.
\end{abstract}

\maketitle

The future success of quantum technologies will depend on the ability to integrate components of different systems. Strongly interacting systems, such as ions \cite{blatt_cnot,wineland_ion} or superconductors \cite{ccqed_gate}, are ideal for processing; large ensembles for memory \cite{si_memory}; and optical photons for communication \cite{cryptography}.
However, interfacing these components remains a challenge. For example, although cavity QED in the microwave domain, using Rydberg atoms  \cite{raimond} or superconducting circuits \cite{wallraff}, provides efficient coupling between photons and static qubits, microwave photons are not ideal for quantum communication due to the blackbody background.
For this reason, quantum interfaces that combine different functions of a network are desirable.

Here we demonstrate a system that allows processing of optical photons using microwave fields \cite{petro08}. We store optical photons in highly excited collective states (Rydberg polaritons) of a cold atomic ensemble using electromagnetically induced transparency (EIT) \cite{eit_review,fleisch00}. Due to the strong dipole-dipole interaction between Rydberg excitations only one excitation is allowed within a volume known as the blockade sphere. Consequently an ensemble smaller than the blockade sphere produces an efficient single photon source \cite{dudin}. Similarly, Rydberg EIT \cite{moha07} gives rise to giant optical non-linearities \cite{pritchard10,Parigi} that can be exploited to modify light at the single photon level \cite{peyronel}.  Here we exploit Rydberg EIT to write a bounded number of photons into a cold atomic ensemble. Subsequently we perform quantum state control of the stored photons using a microwave field resonant with a close--lying Rydberg state. We show that the microwave field modifies the long-range interactions between the stored photons providing a key step toward the realisation of an all--optical analogue of neutral atom quantum gates based on dipole blockade \cite{isen10,wilk10}.

\begin{figure*}[!hbt]
\begin{center}
\includegraphics[width=17cm]{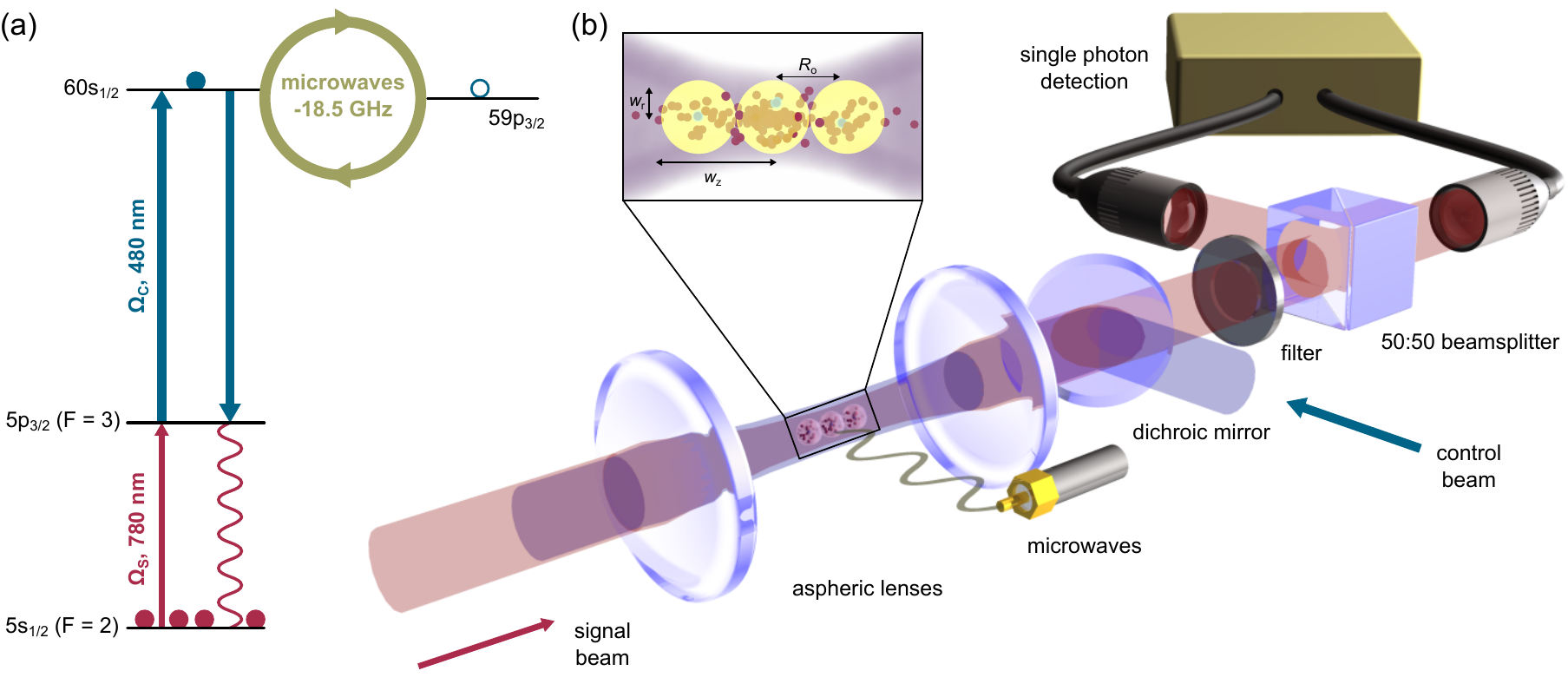}
\caption{Schematic of the experiment. (a) The atomic level scheme used to interface optical photons to the microwave field. (b) Optical photons are stored as Rydberg polaritons in a cold atomic ensemble. Subsequently, the internal states of the polaritons and their interactions are controlled using a microwave field. Finally, the modified optical field is read out and detected using time-resolved single photon counters arranged in a Hanbury Brown-Twiss configuration. }
\label{fig:1}
\end{center}
\end{figure*}

A schematic of the experiment is shown in Fig.\ 1.  Further details are given in the supplemental materials \cite{supp}. An optical dipole trap confines an ellipsoidal atomic cloud containing up to 100 atoms. The approximate axial and radial dimensions of the atomic cloud are \mbox{$w_z=30~\mu$m} and $w_{\rm r}=2.8~\mu$m, where $w$ denotes the standard deviation of the density distribution.  The signal photons, resonant with the $5\textrm{s}^2S_{1/2}(F=2)\rightarrow5\textrm{p}^2P_{3/2}(F=3)$ transition
 in $^{87}$Rb at 780.2~nm (see Fig.\ 1(a)), propagate along $z$. The signal beam is focused to a $1/{\rm e}^2$ radius of $1.2\pm0.1~\mu\textrm{m}$ at the centre of the atomic ensemble. 
A counter-propagating control beam with wavelength $480~\textrm{nm}$ couples the signal transition to a highly excited Rydberg state with principal quantum number $n=60$. The control beam is focused to a $1/{\rm e}^2$ radius of $17.9\pm0.3~\mu$m.  The peak value of the control and signal beam Rabi frequencies are $\Omega_{\textrm{c}}/2\pi=3~\textrm{MHz}$ and $\Omega_{\textrm{s}}/2\pi= 1.2~\textrm{MHz}$, respectively.

In Fig.\ 2(a) we illustrate the photon storage and retrieval process. The signal pulse is stored by reducing the intensity of the control field over a time of 100~ns. At this time optical photons from the signal field are stored as Rydberg polaritons. A microwave pulse then couples the initial Rydberg state to a neighboring Rydberg state (see Fig.\ 1(a)). After the desired storage time, the control field is turned back on to read out the polariton field. This cycle is repeated every 6~$\mu$s. The retrieved signal is typically around 200~ns long, which is determined by the control field switching time. The corresponding bandwidth of the storage process, $\Delta_{\textrm{s}}$, is $1.34\pm0.04~\textrm{MHz}$ (FWHM). We note that the storage efficiency is far from optimised. Efficiencies approaching 100\% are in principle feasible by mode-matching to the time-reversed single-photon emission process \cite{pedersen}. The signal pulse contains approximately 10 photons on average. There is a peak probability of roughly 4\% of retrieving a photon per store/ retrieve experiment. This value has been corrected for the detection efficiency which is approximately 18\%. It is not possible for us to distinguish between the storage efficiency and retrieval efficiency, although the storage efficiency is probably limited by the optical depth of the atomic cloud which is typically around 1.

\begin{figure}[!hbt]
\begin{center}
\includegraphics[width=8cm]{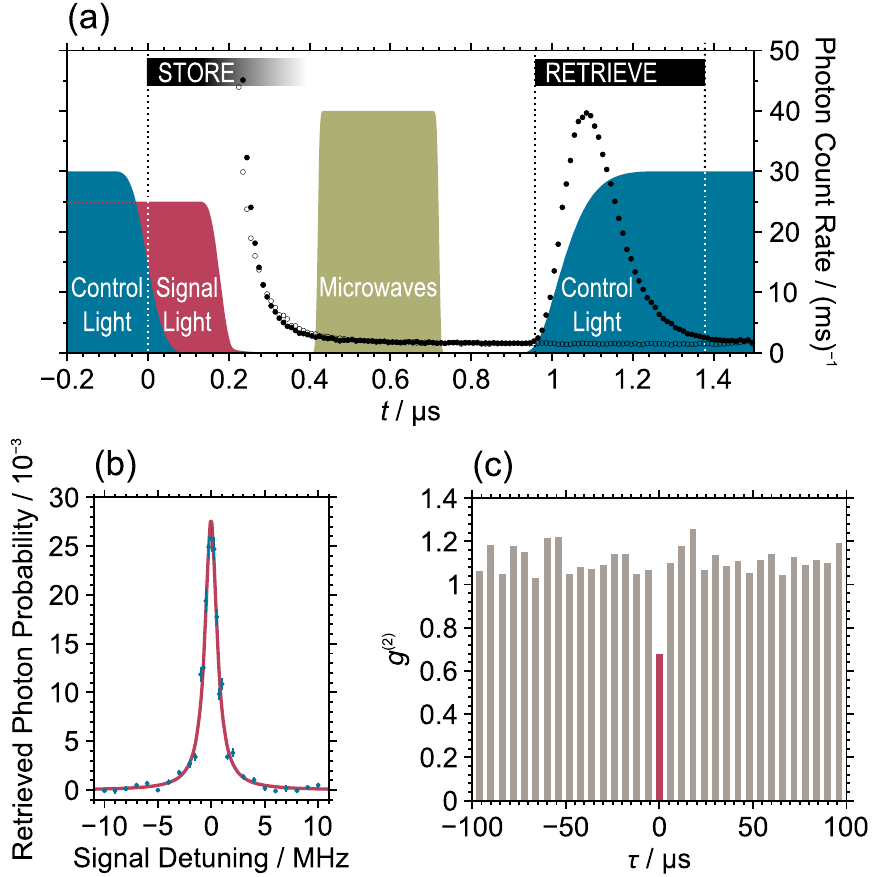}
\caption[]{Photon storage and retrieval: (a) The photon storage process begins at $t=0$ when the control field (blue) is turned off. The signal pulse, which has a total duration (not shown) of $1.1~\mu$s, turns off approximately 200~ns later. After a storage time of roughly $900~\textrm{ns}$ the control field is turned back on to read out the polariton field.  During the storage interval a microwave pulse can be used to couple the polariton to a neighboring Rydberg state. The retrieved signal (filled circles) appears as a peak with a FWHM of $120\pm20~\textrm{ns}$. The background signal without atoms (open circles) is shown for reference. A black band highlights the time window taken as the retrieved signal. The relative heights of pulses are not to scale. (b) The detuning dependence of the retrieved signal indicates the bandwidth of the storage process. The line is a Lorentzian fit with FWHM of $1.34\pm0.04$~MHz. (c) The normalised second-order correlation function $g^{(2)}$ of the retrieved signal, binned over the entire retrieved pulse, as a function of the time delay $\tau$ between the two detectors. The suppression at $\tau=0$ is a signature of dipole blockade during the polariton write process. Note that no microwave coupling has been applied in this case.}
\label{fig:2}
\end{center}
\end{figure}

\begin{figure*}[!hbt]
\begin{center}
\includegraphics[width=17cm]{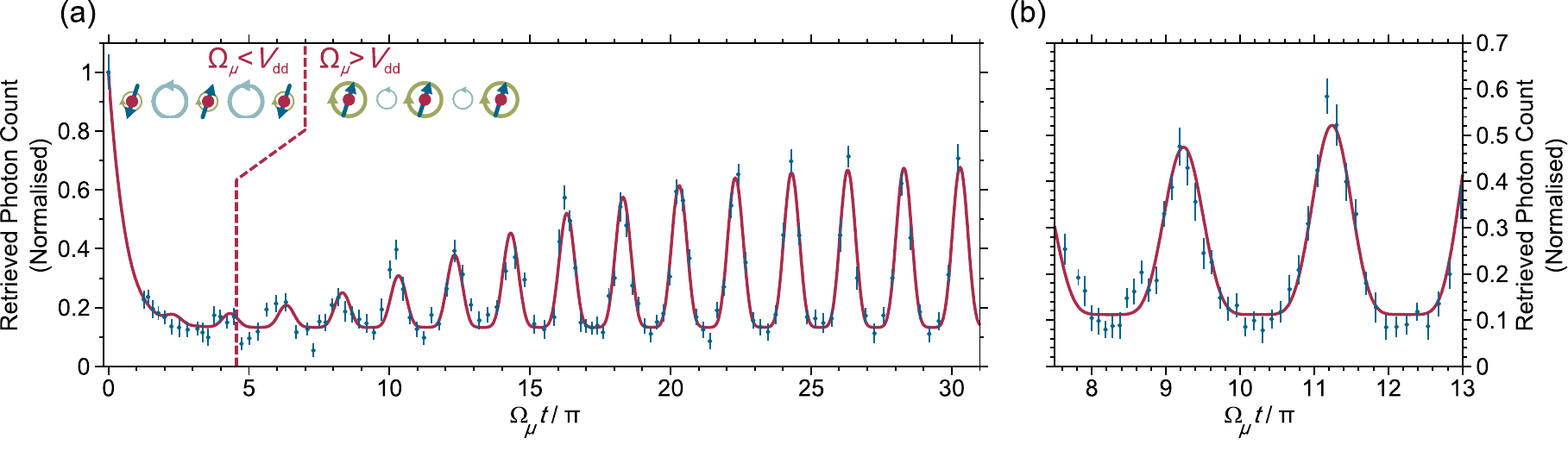}
\caption[]{Controlling the interaction between Rydberg polaritons. (a) The retrieved signal, normalised to the case where no microwave coupling is applied, is plotted as a function of the microwave Rabi frequency, $\Omega_\mu$. The microwave pulse duration is fixed at 300~ns. The dynamics depend on the ratio between the Rabi coupling and the dipole-dipole interaction $\Omega_\mu/V_{\rm dd}$ -- the condition $\Omega_\mu=V_{\rm dd}$ is indicated by the dashed line. For $\Omega_\mu<V_{\rm dd}$ (left hand side), resonant energy exchange between polaritons dominates over Rabi oscillations. For $\Omega_\mu>V_{\rm dd}$, Rabi oscillations dominate and the exchange process is suppressed as the strong driving lifts the degeneracy between the dipole--dipole coupled states.  The solid line is a phenomenological fit using the characteristic form for ${\cal N}$--particle Rabi oscillations coupled to a single optical read out mode, $P=[\cos^{2}(\Omega_\mu t/2)]^{\cal{N}}$. This function is combined with a tanh envelope, and an exponential decay at low microwave Rabi frequencies. From the fit we obtain ${\cal{N}}=2.70\pm0.16$.  Inset: Spin model of the dynamics.  The dipole--dipole interaction (circles between atoms) favours excitation exchange between out-of-phase atomic spins (straight arrows) whereas strong microwave driving (circles around atoms) favours in-phase oscillations and suppresses the exchange process. (b) Higher resolution data of Rabi oscillations in the strong driving regime. The line is a similar fit to Fig. 3(a), with ${\cal{N}}=3.0\pm0.2$. The microwave pulse duration is 150~ns.} \label{fig:3}
\end{center}
\end{figure*}

Let us first consider the situation where no microwave field is applied during the storage interval. Dipole blockade limits the number of excitations that can be written into the sample. The dipole-dipole interaction between Rydberg atoms requires that the polaritons are separated by a distance $R\geq R_{\rm o}=(C_6/\hbar\Delta_{\rm EIT})^{1/6}$ \cite{lukin01}, known as the ``blockade radius'' for optical excitation, where $C_6$ is the van der Waals coefficient that scales as $n^{11}$ and $\Delta_{\rm EIT}$ is the EIT linewidth. For our experimental parameters, the EIT width is $\Delta_{\rm EIT}/2\pi=1~\textrm{MHz}$ and the blockade radius is $R_{\rm o}\approx7~\mu$m. Dipole blockade leads to anti-bunching in the read out pulse as demonstrated in \cite{dudin}. To observe this photon blockade effect, we perform a Hanbury Brown--Twiss coincidence measurement on the retrieved photons (see Fig.\ 1(b)). The photon coincidences characterized by the second-order correlation $g^{(2)}(\tau)$ is plotted in Fig.\ 2(c). There is a peak every $6~\mu$s corresponding to the repetition rate of the experiment. In the absence of photon interactions, the height of the peaks is expected to be unity indicating no bunching or anti-bunching (in practice, variation in storage efficiency throughout an experimental run leads to a level $1.088\pm0.003$).  In contrast the probability of coincidences within each pulse gives $g^{(2)}(0)=0.68\pm0.04$. This partial suppression of $g^{(2)}(0)$ is consistent with a sample that is longer than the blockade radius as shown in Fig.~1(b). The non-zero background signal apparent in Fig.\ 2(a) degrades the measured contrast of $g^{(2)}$ \cite{Orrit}, suggesting that $g^{(2)}(0)$ is about 0.06 lower than observed.

Each photon is stored in the collective polariton state
\begin{eqnarray}
\vert s\rangle=\frac{1}{\sqrt{N}}\sum_{j=1}^N {\rm e}^{{\rm i}\phi_j}\vert s^{j}\rangle~,
\end{eqnarray}
where $\vert s^{j}\rangle=\vert 0_1 0_2\cdots s_{j}\cdots 0_{N}\rangle$, and $N$ is the number of atoms per blockade sphere  \cite{lukin01}. The phase factors are given by $\phi_{j}=\vec{k}\cdot \vec{r_j}$, where $k$ is the effective wavevector of the spin wave, and $r_j$ is the position of atom $j$. The phase of each term in the superposition ensures that the read-out emits a photon into the same spatial mode as the input. The lifetime of this phase-matched polariton is limited to roughly 2~$\mu$s by motional dephasing \cite{dudin}. If atomic motion were reduced by additional cooling, the decoherence time would be ultimately limited by Rydberg lifetime which scales as $n^3$.  For $60{\rm{s_{1/2}}}$ the Rydberg lifetime is of order $100~\mu\rm{s}$. 

We now consider the case where a microwave field is applied during the storage interval (see Fig.\ 2(a)). Coherent control of the stored photon is performed using a resonant microwave field to couple the initial collective state $\vert s\rangle$ to a collective state
\begin{eqnarray}
\vert p\rangle=\frac{1}{\sqrt{N}}\sum_{j=1}^N {\rm e}^{{\rm i}\phi_j}\vert p^{j}\rangle~,
\end{eqnarray}
where $p$ denotes an $n$p Rydberg excitation. The states $\vert s\rangle$ and $\vert p\rangle$ form a two-level basis for collective encoding of the stored photon \cite{pedersen}.  As the dipole moment for the $n{\rm s}\rightarrow n{\rm p}$ transition scales as $n^2$, the figure-of-merit for single qubit rotations (Rabi frequency $\times$ dephasing time) scales as $n^5$. For $n=60$, of order 1000 qubit rotations within the decoherence time are possible. In this collective basis, both the Rabi oscillation frequency and the dephasing rate are independent of the atom number $N$ (in contrast to the transition from the ground state $\vert 0_1\cdots0_N\rangle$  to the collective state $\vert s\rangle$, where the Rabi frequency scales as $\sqrt{N}$ \cite{heidemann07}). This is important, as it allows us to observe the high contrast oscillations over many cycles even for a non-deterministically loaded sample.

We study Rabi oscillations for $n=60$ polaritons coupled to a microwave field resonant with the 60s$_{1/2}\rightarrow$59p$_{3/2}$ transition at 18.5~GHz. As the read out state is $\vert s\rangle$, the retrieved signal oscillates between a maximum when the polariton state is $\vert s\rangle$ and a minimum when the polariton state is $\vert p\rangle$. In Fig.\ 3 we plot the retrieved photon signal as a function of the rotation angle, $\Theta=\Omega_\mu t$. The microwave pulse duration is fixed so the microwave Rabi frequency $\Omega_\mu$ increases from left to right in Fig.\ 3. Counter-intuitively, the Rabi oscillations revive for large $\Theta$.

To understand these unusual dynamics, consider the pairwise dipole--dipole interaction between the cycling Rydberg polaritons \cite{supp}. The microwave coupling between $\vert s\rangle$ and $\vert p\rangle$ induces resonant dipole--dipole interactions between polariton modes with an interaction energy $V_{\rm dd}=~d^2/(4\pi \epsilon_0 R_{\rm o}^3)$, where $R_{\rm o}$ is the correlation length associated with the 60s blockade process. The microwave field thus introduces a second blockade scale \cite{monsit,brekke12} with a characteristic size $R_{\mu}=(C_3/\hbar\Omega_\mu)^{1/3}$, where $C_3$ is the resonant dipole-dipole interaction coefficient and $\Omega_\mu$ is the Rabi frequency of the microwave transition. By varying $\Omega_\mu$ we can tune the ratio $R_\mu/ R_{\rm o}$. For the range of Rabi frequencies shown in Fig.~3(a) we change between a regime where $\Omega_\mu<V_{\rm dd}$ on the left hand side and  $\Omega_\mu>V_{\rm dd}$ on the right hand side.  

For $\Omega_\mu<V_{\rm dd}$, the resonant dipole-dipole interaction associated with the microwave transition dominates. In this case, the blockade sphere associated with the microwave transition is larger than the blockade sphere associated with the formation of the 60s polaritons, $R_\mu> R_{\rm o}$. 
As the resonant dipole--dipole interaction is an exchange process \cite{ates,ditz}, this regime is dominated by excitations hopping, leading to loss or dephasing of the polariton read-out \cite{bariani12}. 
Consequently the retrieved photon signal is suppressed and fits to an exponential decay. In this dephasing regime one may expect only a single excitation to survive and hence strong anti-bunching in the retrieved photon signal \cite{bariani12}. This effect was not observable in the current experiment, as the background signal apparent in Fig.\ 2(a) contributes a larger fraction of the retrieved signal in the dephasing regime. The data in Fig.\ 3 was also acquired under less well optimised conditions, where the background signal was up to 50\% of the signal corresponding to the peaks of the suppressed Rabi oscillations. 

For $\Omega_\mu>V_{\rm dd}$ strong driving forces the dipoles to oscillate in phase, which suppresses the out-of-phase exchange interaction. In this case, where the microwave blockade radius is smaller than the optical blockade, $R_\mu< R_{\rm o}$, the exchange or hopping term is reduced to $V_{\rm dd}^2/\Omega_\mu$.  Consequently the spin wave dephasing is reduced and the Rabi oscillations reappear. This recovery in the Rabi oscillations is a direct signature of the spatial correlations between Rydberg polaritons, and occurs when the microwave Rabi frequency is sufficient to overcome the polariton-polariton blockade. We note in passing that Lamor dephasing of the spin wave \cite{mat06} is not expected to have a significant effect on the dynamics of the system, since $\Omega_\mu$ is in general much larger than the Lamor frequency.

Significantly, ${\cal N}$-particle correlations in the read out give rise to enhanced sensitivity to the rotation angle $\Theta$, which could be exploited in quantum metrology applications \cite{duan11}. The retrieval probability, 
\begin{eqnarray}
P=\left[\cos^{2}\left(\frac{\Theta}{2}\right)\right]^{\cal{N}}~,
\end{eqnarray}
is given by applying a Wigner rotation matrix to the collective Dicke state of ${\cal N}$ spins \cite{supp}. This many-body character of the collective read-out is clearly visible in Fig.\ 3(b). If we fit to $[\cos^{2}(\Omega_\mu t/2)]^{\cal{N}}$ allowing $\cal{N}$  to float we obtain ${\cal{N}}=3.0\pm0.2$. This is consistent with the number of blockade spheres in our ensemble, given the 60s blockade radius and the geometry of the atomic cloud (see Fig. 1(b)).

 In conclusion, we have demonstrated control over the quantum state of Rydberg polaritons using a microwave field. By tuning the strength of the microwave field we have shown that the interaction between neighboring polaritons can be varied. This effect was observed in Rabi oscillations of the polariton state, which exhibit a many-body character consistent with ${\cal{N}}=3$ Rydberg excitations. The ability to control the quantum state of Rydberg polaritons opens some interesting prospects for advances in quantum information and quantum simulation of strongly correlated systems. For example, the competition between resonant energy exchange (hopping) and localisation is reminiscent of the Jaynes-Cummings-Hubbard model \cite{greentree,supp}. In addition, Rydberg polaritons provide a powerful platform for studying strongly coupled atom--light interactions without a cavity, quantum simulation of spin liquids \cite{lesan12}, and quantum metrology using Dicke states \cite{duan11}.
The ability to control the interactions between polaritons using microwave fields allows a second blockade scale to be established. This provides a viable route towards fully deterministic photonic phase gates using single photons \cite{peyronel}, or to generate non-classical states of light from classical input fields \cite{stano12}. It is also an ideal system to study resonant energy transfer \cite{ates}.
Finally, Rydberg polaritons provide a convenient interface between quantum systems that operate in the microwave and optical domains, such as circuit QED \cite{wallraff} and atomic ensembles, respectively. Rydberg polaritons act as a source of quantum light, that can be coupled to on-chip \cite{nirrengarten} microwave resonators which in turn interface to solid state qubits \cite{petro08,wallraff}, forming a complete architecture for transmitting, storing and processing quantum information.

We acknowledge financial support from Durham University, EPSRC and the EU Marie Curie ITN COHERENCE Network. We thank C. Ates and T. Vanderbruggen for stimulating discussions, and are grateful to A. West for assistance with the figures. We also thank the referees for their useful comments.
\\
\\
\hyperref[con1]{*}daniel.maxwell@durham.ac.uk
\hyperref[con2]{*}c.s.adams@durham.ac.uk

\onecolumngrid  
\section*{Supplemental Materials}

\subsection{Further experimental details}

Laser cooled $^{87}{\rm Rb}$ atoms are loaded into an optical dipole trap at $\lambda=910~\textrm{nm}$. The trapping light, which co-propagates with the signal beam, is focussed to a $1/{\rm e}^{2}$ radius of approximately $5~\mu{\rm m}$. The trap depth is roughly $0.6~{\rm mK}$ and the temperature of the trapped atomic cloud is typically $\approx100~\mu\textrm{K}$. After repumping the atoms into the $5\textrm{s}_{1/2}(F=2,m_F=2)$ state, the dipole trap power is switched on-and-off with a period of $6~\mu \textrm{s}$ and a 50\% duty cycle. Experiments are performed during the periods when the trap is off to avoid differential light shifts, with one store/retrieve experiment per trap-off period. The trap modulation typically lasts $20~\textrm{ms}$, after which the storage efficiency begins to decrease. The trap is then reloaded and the sequence is repeated.

When the dipole trap has been switched off, a signal pulse of $1.1~\mu\textrm{s}$ duration is stored by reducing the intensity of the control field. A microwave pulse of duration $150-300~\textrm{ns}$ is applied, coupling the $60\textrm{s}_{1/2}$ and $59\textrm{p}_{3/2}$ Rydberg states. The dipole moment for this transition is $d=\sqrt{2/9}\times3468~ea_0$, where the radial matrix element was calculated using the Numerov method. The microwaves are emitted from a stub antenna, orientated such that the microwaves drive $\pi$-transitions. Following the microwave pulse, the control field is turned back on to read out the polariton field. Rabi oscillations are mapped out by varying the power of the microwave pulse, keeping the pulse length fixed. The total storage time of the polariton is approximately $900~\textrm{ns}$.

The retrieved photon signal is windowed over the region highlighted in Fig.\ 2(a). When calculating $g^{(2)}$, only photons within this time window are correlated. For the Rabi oscillation measurements in Fig.\ 3, the total number of photon counts within the window is extracted for each shot. There are 30 shots for each data point, with 3334 individual experiments in each shot.  

\subsection{Collective Rabi oscillations of ${\cal N}$ indistinguishable Rydberg polaritons}  

Photon storage in the regime of dipole blockade forms the polariton superposition state 
\begin{eqnarray}
\vert s \rangle & = & \frac{1}{\sqrt{N}} \sum_{j=1}^N  e^{i \phi_j} \vert s^{j} \rangle~,
\end{eqnarray}
where $\vert s^{j} \rangle = \vert 0_1 0_2\ldots s_j \ldots 0_N \rangle$ is a state with the single excitation at atom $j$ within an ensemble of $N$ atoms. The spatial dependence of the phase factors $\phi_j=\vec{k}\cdot \vec{r_j}$, where $\vec{k}=\vec{k}_s+\vec{k}_c$, is determined by the sum of the wavevectors of the excitation lasers and the position of each atom $\vec{r}_j$. The phase factors determine the directionality of the optical read out of the polariton.

A resonant microwave field couples the singly-excited, symmetric polariton state $\vert s\rangle$ to a similar state
\begin{eqnarray}
\vert p \rangle & = & \frac{1}{\sqrt{N}} \sum_{j=1}^N  e^{i \phi_j} \vert p^{j} \rangle~.
\end{eqnarray}
Under microwave coupling each Rydberg polariton (neglecting polariton-polariton interactions) becomes an effective two-level system or spin-$\textstyle{1\over2}$ quasi-particle. As the wavelength of the microwaves is much larger than the sample size, the global phase-structure of the polariton is preserved. Any operator acting on this system can be written in terms of the Pauli spin matrices $\sigma^x$, $\sigma^y$, and $\sigma^z$. For ${\cal N}$ polaritons we define collective spin operators
\begin{eqnarray}
\mbox{\boldmath$J$}_x = \frac{1}{2}\sum_{j=1}^{\cal N}\sigma^x_j~,
\end{eqnarray}
and similarly for $y$ and $z$. The Dicke states labelled $\vert J,M\rangle$ are the eigenstates of $\mbox{\boldmath$J$}^2 = \mbox{\boldmath$J$}_x^2+\mbox{\boldmath$J$}_y^2+\mbox{\boldmath$J$}_z^2$ and $\mbox{\boldmath$J$}_z$.
For coherent driving, the evolution of the ${\cal N}$-particle Dicke states is given by the Wigner rotation matrix $ {\cal D}$. The reduced rotation matrix between states with the same $J$ but projections $M'$ and $M$ is \cite{rose}
 \begin{eqnarray}
{\cal D}^J_{M'\!\!,M}(0,\Theta,0)& =&\frac{(-1)^{M' - M}}{(M' - M)!} \sqrt{\frac{ (J-M)!(J+M')!}{(J+M)!(J-M')!}}\cos^{2J+M -M'}\left(\frac{\Theta}{2} \right) \sin^{M'-M}\left(\frac{\Theta}{2} \right) F~,
 \end{eqnarray}
where $F=\,_2F_1(M'-J, -M-J; M'-M+1; -\tan^2\Theta/2)$ is a Gauss hypergeometric function, and the angle $\Theta = \Omega_\mu t$ is the rotation induced by the microwave field.
The initial spin state is $\vert J={\cal N}/2,M\!=\!-{\cal N}/2\rangle$. 
To retrieve the stored optical pulse in the same mode we need to conserve the phase-structure, therefore we project onto the identical state $\vert J={\cal N}/2,{M}^\prime\!=\!-{\cal N}/2\rangle$. In this particular case, where $M=-J$, the hypergeometric function takes the value 1. The probability of photon retrieval is proportional to the matrix element squared giving 
\begin{equation}
P = \left\vert {\cal D}^{{\cal N}/2}_{{\cal N}/2,{\cal N}/2} \right\vert^2 = \left[\cos^{2}\left(\frac{\Theta}{2} \right)\right]^{{\cal N}}~. 
\end{equation}
This result gives an accurate description of the collective read out of the rotation of the ${\cal N}$-polariton state. Note that the frequency of the oscillations corresponds to the single-atom Rabi frequency $\Omega_\mu$, and  does not depend on the number of polaritons ${\cal N}$, or the number of atoms in each polariton $N$ (assuming that $N$ is sufficiently large to ensure directional emission). This feature is important, since it allows us to observe the oscillations in a non-deterministically loaded sample. The power of the cosine depends only on the mean number of excitations ${\cal N}$ arising during to the photon storage protocol. Larger ${\cal N}$ gives enhanced sensitivity to the rotation angle $\Theta$ around integer multiples of $2\pi$.

Finally, it is  important to point out that this picture only applies if the number of atoms, $N$, in each polariton, and their spatial extent are sufficiently large to support directional emission during read out. The condition on the size is that the blockade radius should be many times the wavelength of the optical transition $R_{\rm o}\gg \lambda$.  Both the size and number conditions break down at low principal quantum number, $n$, and the dynamics revert to single particle Rabi oscillations.

\subsection{Spin chain of strongly interacting polaritons}

If we treat the Rydberg polaritons as localised spin-$\textstyle{1\over 2}$ quasi-particles with a separation $R_{\rm o}$, then the interaction Hamiltonian for zero detuning and only nearest-neighbour interactions is
\begin{eqnarray}
H&=& g\sum_{i=1}^N  (\sigma_i^+ a_i+ \sigma^-_i a^\dag_i)+V_{\rm dd}\sum_{\langle ij \rangle}\sigma^+_i\sigma^-_j+f\sum_{i=1}^N  (a_i+ a^\dag_i)~,
\label{eq:spins}
\end{eqnarray}
where the first term is the usual Jaynes-Cummings coupling between a two level system and a quantised microwave field, the second term with $V_{\rm dd}=d^2/4\pi \epsilon_0 R_{\rm o}^3$ describes the dipole--dipole interaction between sites $i$ and $j$, and the third term describes coherent driving. We note in passing the analogy to the Jaynes-Cummings Hubbard Model (JCHM) which describes photon tunnelling between arrays of cavities with one two-level system in each cavity \cite{nissen12}.  
The main difference is that whereas in the JCHM real photons hop between cavities, for Rydberg polaritons there is an exchange of virtual photons, i.e., the photon hopping term $a^\dag_ia_j$ in the JCHM  is replaced by an excitation hopping term $\sigma^+_i\sigma^-_j$.

The current experiment differs from the situation described by equation (\ref{eq:spins}) in a number of ways. First, the absence of a microwave cavity to enhance the single photon coupling $g$ means that we can treat the microwave field classically. Second, the field drives a $\pi$-polarized s$_{1/2}\rightarrow$ p$_{3/2}$ transition but the dipole-dipole interaction can also couple to other angular momentum states. Neglecting the fine and hyperfine structure, the interaction Hamiltonian has the form
\begin{eqnarray}
H&=& \frac{\Omega_\mu}{2} \sum_{i=1}^N  \mu_i^z+V_{\rm dd} \sum_{\langle ij \rangle}(\mu^+_i \mu^-_j +\mu^-_i \mu^+_j- 2\mu^z_i\mu^z_j)~,
\label{eq:polaritons}
\end{eqnarray}
where $\Omega_\mu$ is the Rabi frequency on the microwave transition and $\mu_i^{z,\pm}$ is the dipole operator for $\Delta m=0,\pm1$ transitions on the $i$-th polariton. The eigenvalues for two localised polaritons as a function of their separation are shown in Fig.~4. The energy scales can be parameterized in terms of the two radii, $R_{\rm o}=(C_6/\hbar\Delta_{\rm EIT})^{1/6}$ and $R_{\mu}=(C_3/\hbar\Omega_\mu)^{1/3}$, characterising the optical and microwave blockade, respectively. For 60s60s $C_6=-140$~GHz~$\mu{\rm m}^6$ and for 60s59p$_{3/2}$ $C_3=-14.3$~GHz~$\mu{\rm m}^3$ \cite{singer,jon}.  For weak driving, Fig. 4(a), there is a strong state mixing at $R_{\rm o}$ which suppresses the retrieved photon signal whereas for strong driving, Fig. 4(b), the dressed states are only weakly perturbed by dipole-dipole interactions.

A final consideration is that the polaritons are delocalised. Consequently, the dipole--dipole exchange terms are different for each atom pair, $\mu,\nu$, and we should replace $V_{\rm dd}$ with a sum over individual atoms in each polariton $V_{\rm dd}^{(\mu,\nu)}$.
In the regime, $R_\mu >R_{\rm o}$, this gives rise to loss and dephasing of the spin wave \cite{bariani12} as illustrated in Fig.\ 3(a).

\begin{figure}[!hbt]
\begin{center}
\includegraphics[width=17cm]{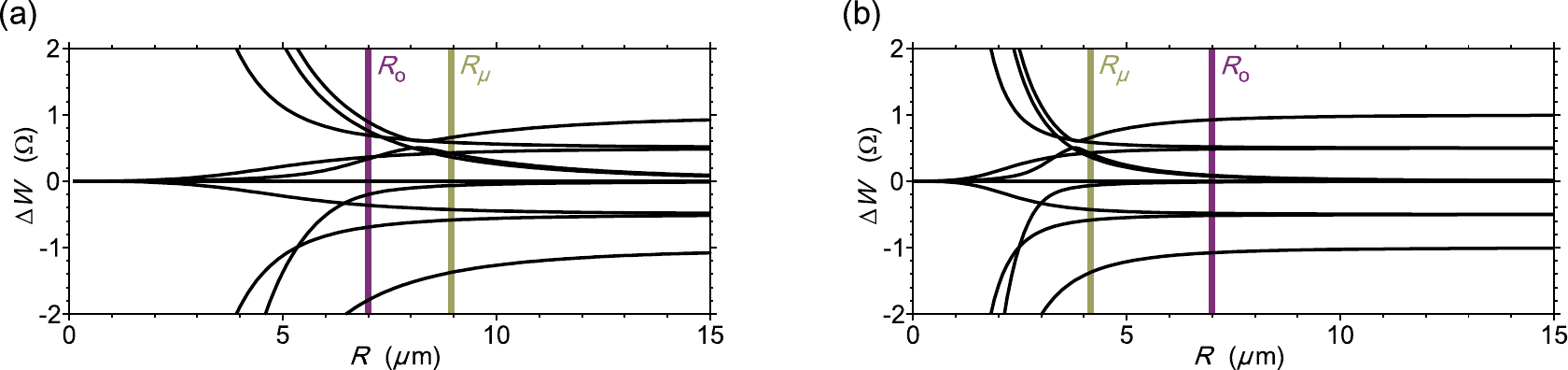}
\caption{Eigenvalues of the polariton Hamiltonian (\ref{eq:polaritons}) as a function of the polariton spacing. The blockade radii for optical, $R_{\rm o}=7~\mu$m, and microwave, $R_\mu$, excitation are indicated. (a) For weak driving $\Omega_\mu/2\pi=20~$MHz, the dipole-dipole interaction dominates, $R_\mu >R_{\rm o}$, and there is a strong mixing of states with different angular momentum, $m$. (b) In contrast for strong driving $\Omega_\mu/2\pi=200~$MHz, $R_\mu <R_{\rm o}$, the splittings between the coupled dressed states are only weakly perturbed by the dipole-dipole interaction (of order $V_{\rm dd}^2/\Omega_\mu$) and there are no level crossings for $R\geq R_{\rm o}$.}
\label{fig:S1}
\end{center}
\end{figure}

\end{document}